\documentclass[11pt,dvips]{article}
\usepackage{epsfig,times}
\usepackage{picinpar}
\makeatletter
\renewcommand{\section}{\@startsection{section}{1}{0in}
        {0.4\baselineskip}{0.1\baselineskip}{\Large\bf}}
\renewcommand{\subsection}{\@startsection{subsection}{2}{0in}
        {0.25\baselineskip}{-\baselineskip}{\large\bf}}
\renewcommand{\subsubsection}{\@startsection{subsubsection}{3}{0in}
        {0.1\baselineskip}{-\baselineskip}{\normalsize\bf}}

\makeatother
\newcommand{\etal}{{\it et al.}}
\begin{document}
\thispagestyle{myheadings}
\markright{OG 2.1.39}
\begin{center}
{\LARGE \bf Short-timescale Variability in the Broadband 
Emission of the Blazars Mkn421 and Mkn501}
\end{center}

\begin{center}
{\bf M.J. Carson$^{1}$, B. McKernan$^{1}$, T.Yaqoob$^{2}$, D.J. Fegan$^{1}$}\\
{\it $^{1}$Department of Physics, University College Dublin, Dublin 4,
Ireland\\ $^{2}$Laboratory for High Energy Astrophysics, Code 660.2
Goddard Space Flight Center, Greenbelt, MD 20771}
\end{center}

\begin{center}
{\large \bf Abstract}
\end{center}
\vspace{-0.5ex}
We analyse \emph{ASCA} x-ray data and Whipple $\gamma$-ray data from the
blazars Mkn421 and Mkn501 for short-timescale variability. We find no
evidence for statistically significant ($>3\sigma$) variability in
these data, in either source, on timescales of less than $\sim 10$
minutes. 
\vspace{1ex}
\section{Introduction}
The recently detected TeV radiation from the Blazars Mkn421 and Mkn501 may be
Comptonized synchrotron radiation (Comastri \etal, 1997) or Comptonized ambient
radiation (Sikora \etal, 1997) or some mixture of the two.  Alternatively, the
$\gamma$-radiation may result from a proton-initiated cascade (Beall \& Bednarek,
1998). Models of the $\gamma$-radiation preict large qualitative differences in some
source parameters, so a search for short-timescales of variability in these AGN
may be important in constraining models of emission and ultimately in
distinguishing between them. Short timescale variability may also constrain
quantum gravity effects predicted by some string theories
(Amelino-Camelia \etal, 1998).\\
Variability studies (using the $\chi^{2}$-test) have been carried out on data from
Mkn501 and Mkn421. The shortest observed timescales of variability in TeV
$\gamma$-rays correspond to doubling times in rapid flares of $\sim 15$ minutes for
Mkn421 (McEnery, 1997) and $\sim2$ hours for Mkn501 (Quinn \etal, 1999) which may
indicate a variability cut-off or a lack of instrumental or method sensitivity
to real variability at timescales shorter than these. The shortest timescales so
far observed in these sources at x-ray energies are $\sim$1 day for Mkn501
(Kataoka \etal,
1999) and 0.5-1.0 days for Mkn421 (Takahashi, Madejski \& Kubo, 1999).\\
The analysis reported here involves a broadband variability study of Mkn501 and
Mkn421 using the Excess Pair Fraction (EPF) technique (Yaqoob \etal, 1997). We
analyse non-simultaneous Whipple TeV $\gamma$-ray data and {\it ASCA} x-ray data
from these blazars.
\section{EPF applied to x-ray data}
\begin{figwindow}[1,r,%
{\mbox{\epsfig{file=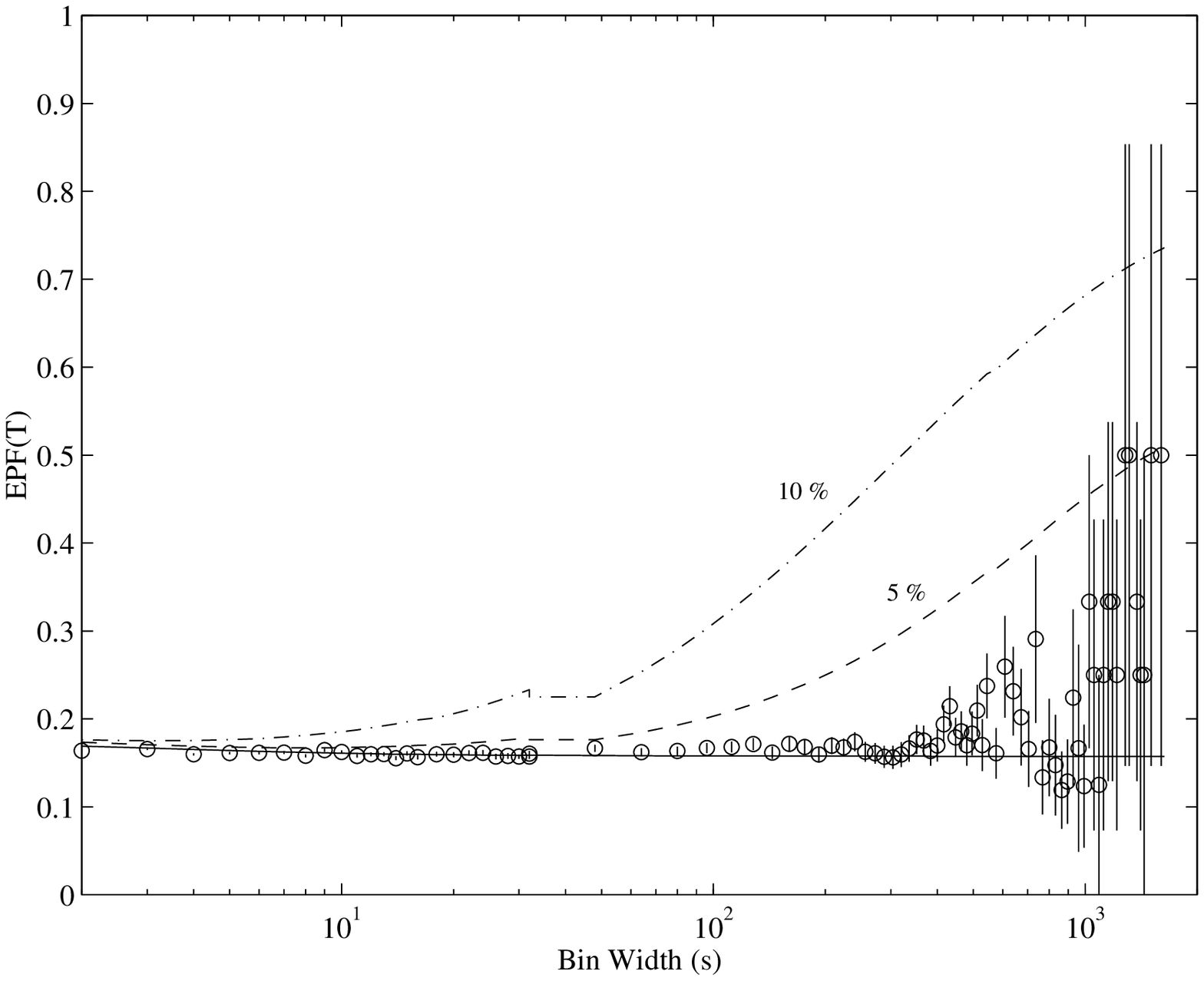,width=8cm}}},%
{EPF for {\it ASCA} x-ray data on Mkn421.}] 
{\it ASCA} (Tanaka \etal 1994) observed Mkn 421 and Mkn 501, obtaining moderate
energy resolution spectra in the 0.5--10 keV band, with a time-resolution of
better than 2 s for two of the four instruments onboard (4 s for the other two).
Mkn 421 was observed on eighteen occassions between 1993, May 10 and 1997, June
3, with a total exposure time of $\sim 240\times10^{3}$s. The 1996 data do not overlap with
the TeV data reported in this paper.  Mkn 501 was observed on four occassions
between 1996, March 21 and 1996, April 2, with a total exposure time of $\sim 49\times10^{3}$
s. Additional {\it ASCA} observations of both sources exist but only those made
of the dates mentioned above were in the public archive at the time of writing.
An excellent account of the Mkn 421 {\it ACSA} observations and results can be
found in Takahashi, Madejski, and Kubo (1999) and the {\it ASCA} results for Mkn
501 can be found in Kataoka \etal  (1999).  We re-analysed the {\it ASCA}
data ourselves, which were reduced in the same manner as described in Yaqoob
\etal (1998). We made EPF for each source following the methods described
in Yaqoob \etal  (1997), averaged over all the available data sets. 
\end{figwindow}
\begin{figwindow}[1,r,%
{\mbox{\epsfig{file=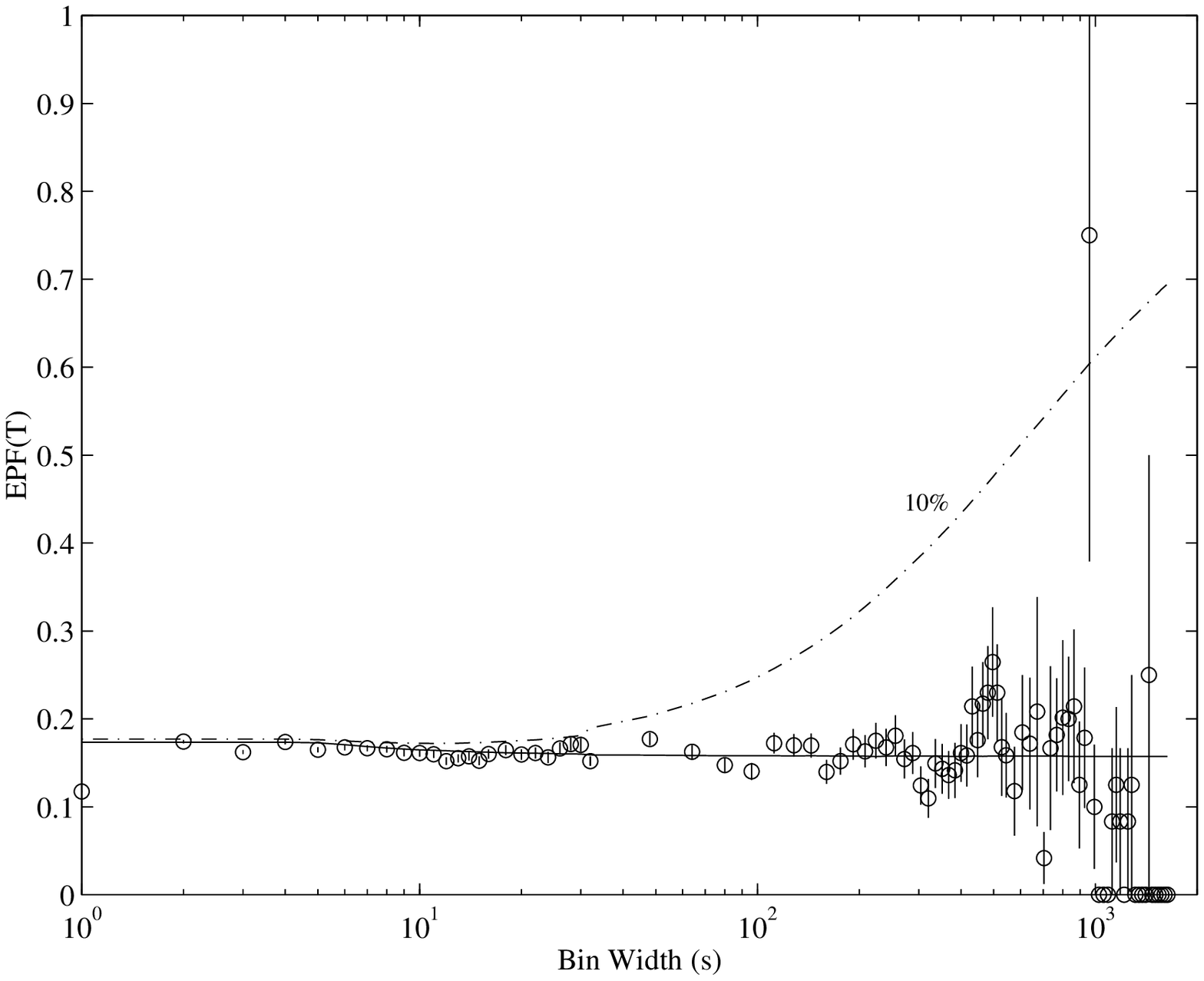,width=8cm}}},%
{Open circles show the Excess Pair Fraction (EPF) computed from all the {\it
    ASCA} Mkn 501 observations (see text).}]  The results for both sources are
shown in Figs. (1) $\&$ (2). The solid, dotted and dashed curves show theoretical
EPF for a source with $0\%$, $5\%$, and $10\%$ random amplitude variability
respectively.  The curves were generated as described in detail in Yaqoob {\it
  et al.} 1997.  It can be seen that in Mkn 421 we can rule out variability at
the $5\%$ level with a confidence level greater than 3$\sigma$ from $\sim500$ s down
to timescales of $\sim 11$ s. (Note: The kink at 32 s is due to only two of the
four {\it ASCA} instruments being used below this timescale because the two
Solid State Spectrometers have less timing resolution than the two Gas Imaging
Proportional Counters). For Mkn 501, for which much less data are available, we
can rule out variability at the $10\%$ level with a confidence level greater
than $3\sigma$ from
$\sim400$ s down to timescales of $\sim 32$ s. \\
\end{figwindow}

\section{EPF applied to $\gamma$ -ray data}
\begin{figwindow}[1,r,%
{\mbox{\epsfig{file=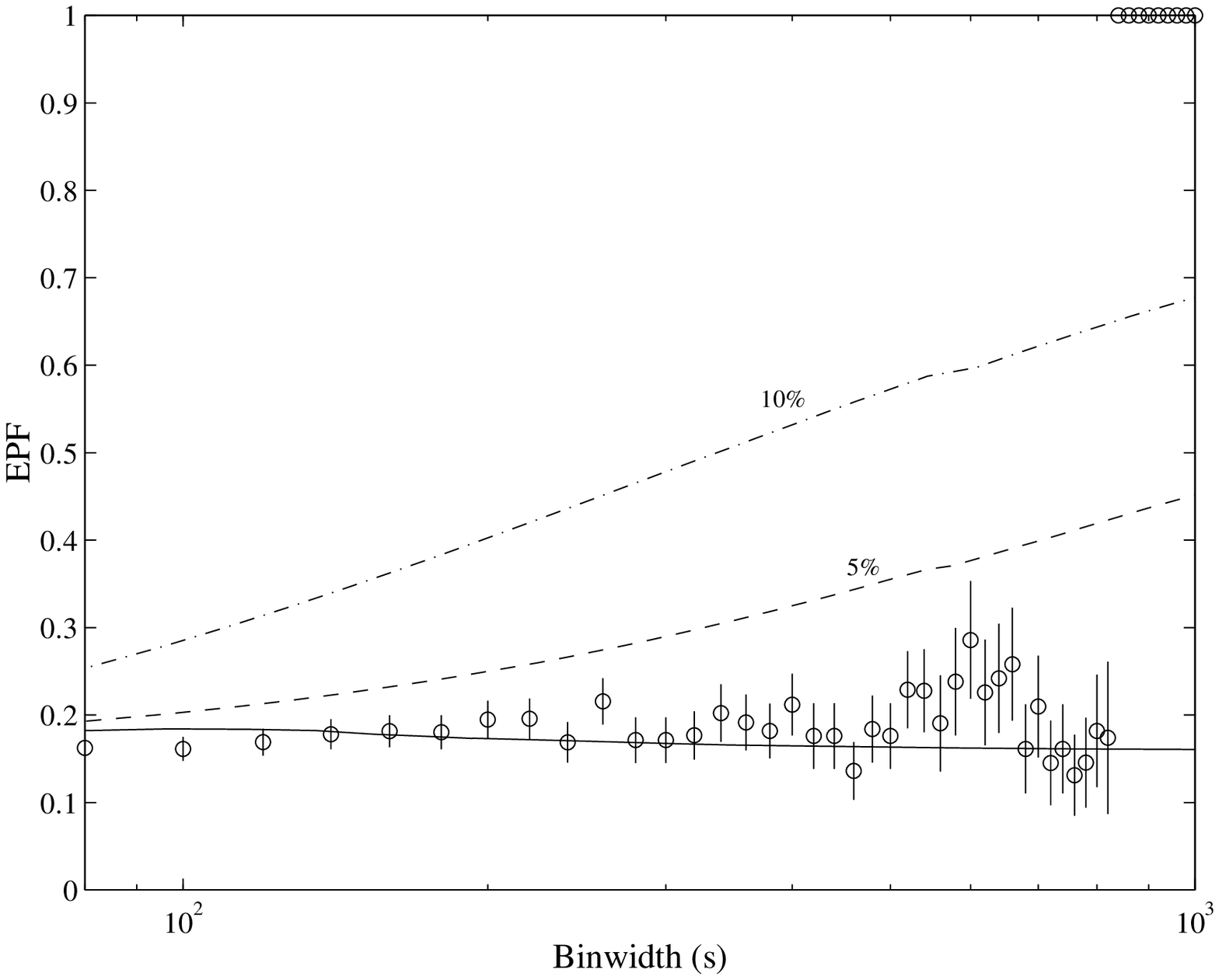,width=8cm}}},%
{Open circles show EPF for Whipple Mkn421 data. The solid line is the
  theoretical EPF expected from a constant source and dashed lines are the
  theoretical EPF for 5\% and 10\% random amplitude variability.}]
TeV data are recorded by the Whipple Observatory Imaging Atmospheric
${\rm \hat{C}}$erenkov Telescope (IACT). The IACT uses the
well-established Extensive Air Shower (EAS) technique to indirectly
observe very high energy $\gamma$-rays from sources such as AGN jets
and supernova remnants (Cawley \etal, 1990). Most atmospheric showers are due
to the collision of high energy cosmic rays (hadrons or leptons),
therefore the parent $\gamma$-ray data, as observed by the IACT, have
a very small signal to noise ratio. Parameter cuts are used to
eliminate most of the noise whilst retaining at least $50\%$ of the
signal. The TeV data are typically recorded in $\sim 28
\rm{minute}$ intervals, at different elevations and during various
weather conditions. Only those data sets with the most stable raw
count rates were used. The Mkn421 data were
taken from the period April 1996-May 1996, during which time the most
significant flaring event ever seen at this energy was observed. The
Mkn501 data were taken from the period April 1997- May 1997, during
which time this AGN was at its most active since its discovery at this
energy in 1995. The presence of strong flaring within these data is
important for statistical purposes as the count rate from the AGN increases
from typically $\leq 1
\gamma\,$min$^{-1}$ to $\sim 10 \gamma\,$min$^{-1}$ during a strong flaring
state. 

\end{figwindow}

\begin{figwindow}[1,r,%
{\mbox{\epsfig{file=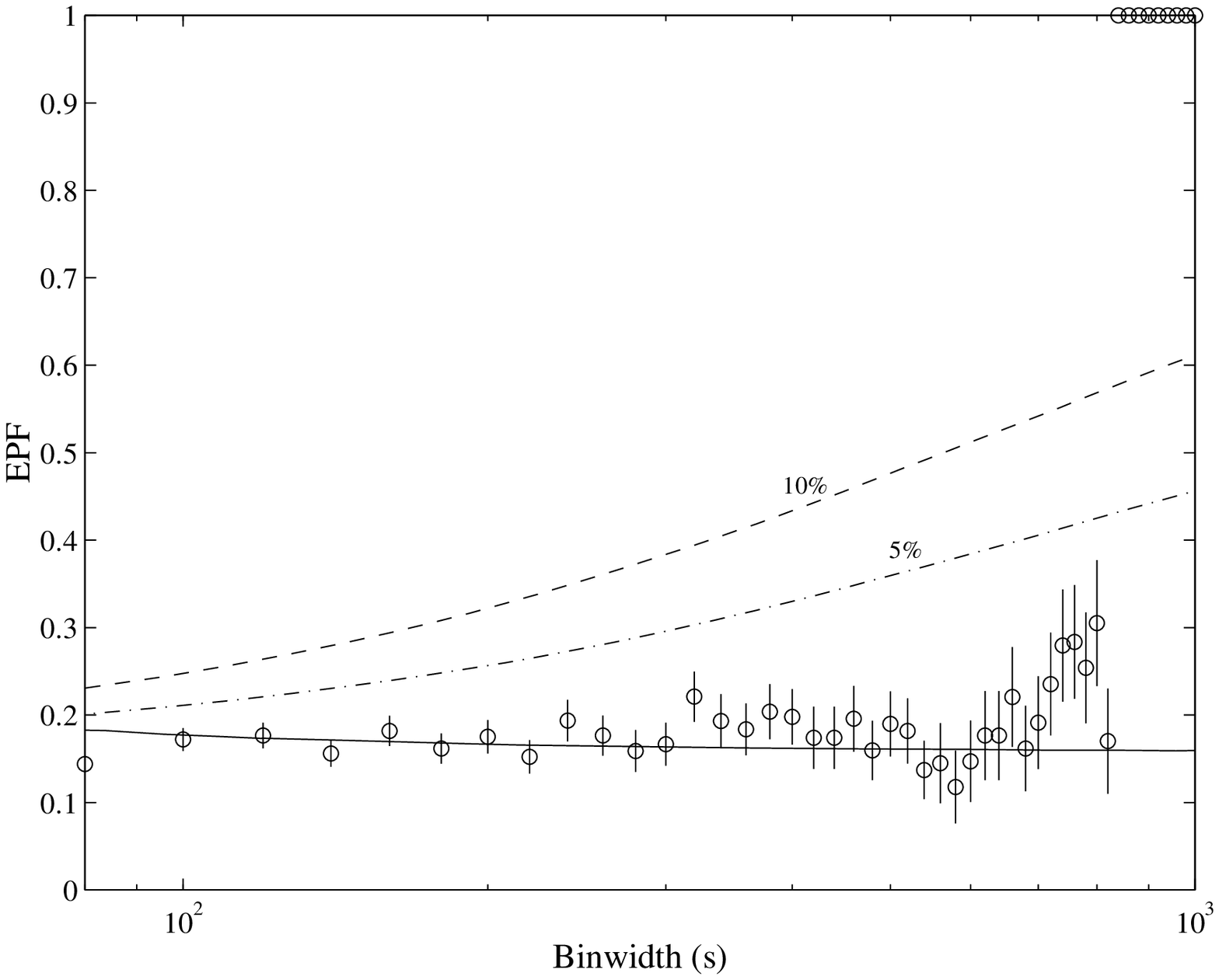,width=8cm}}},%
{EPF for Whipple Mkn501 data.}]  Fig. (3) shows EPF for Mkn421. Approximately
1760 minutes of data are used in this analysis at TeV energies. None of the data
points are greater than $3\sigma$ from that of a theoretical constant source with
the same mean count rate. Because of the 28 minute data binning only timescales
less than 14 minutes can be considered using this technique. The $\gamma$-ray rate
($\sim3\gamma$min$^{-1}$) prevents us probing timescales much lower than $\sim$200
seconds.  Fig. (4) shows EPF for Mkn501 with approximately 1900 minutes of data.
Again none of the data points are greater than $3\sigma$ from that of a constant
source with the same mean count rate indicating a lack of significant
variability in this source below $\sim 10$mins. Here the rate is $\sim9$min$^{-1}$
and so the lower timescale below which the statistics become too poor is $\sim$100
seconds. Thus, we can rule out variability in both sources below $\sim$10 minutes
at a confidence level $>3\sigma$ at the $10\%$ variability level.
\end{figwindow}
\section{Discussion and Conclusions}
Our results thus far show that there appears to be no statistically significant
variability in either of the two AGN, below timescales of $\sim
600$s. From these results we conclude
that either the statistics of the present data are insufficient to
establish significant variability or the data does not vary on these
timescales. If we assume that the source is not varying on timescales shorter
than 10 minutes then we can estimate the minimum size of the emission region:
the Doppler beaming factor is given by $D=1/(1+z)\Gamma_{b}(1-\beta cos\theta)$ where
$z$ is the redshift of the source, $\Gamma_{b}$ is the Lorentz boost, $\beta=v^{\prime}/c$
(where $v^{\prime}$ is the velocity of the emitting region) and $\theta$ is the angle
between the jet axis and the observer. Causal arguments require that the size of 
the emission region $R_{em}$ is constrained to satisfy 
\begin{equation}
R_{em}\,\leq\,c\,\delta t\,\frac{D}{(1+z)} 
\end{equation}
Where $\delta$t is the observed variability timescale ($>10$ minutes here).
For Mkn421 $z=0.031$, $D\approx13.8$ and so $R_{em}\,\approx\,2.5\times10^{14}$ cm. If the
emission region is located in the putative jet then given a jet opening angle of 
$3^{o}$, the emission region is at least $4.7\times10^{15}$ cm from the base of the
jet, or assuming a black hole mass of $\approx10^{8}\,M_{\odot}$, the emission region is
at least $\sim$160 gravitational radii from the black hole.
\section{Acknowledgements}
We would like to thank the Whipple collaboration for the use of their
data. We also thank the HEASARC at NASA/GSFC for use of the {\it ASCA}
public data archive.
\vspace{1ex}
\begin{center}
{\Large\bf References}
\end{center}
Amelino-Camelia, G., \etal 1998, Nature 393, 323\\
Beall, J.H. and Bednarek, W. 1998, astro-ph/9802001\\
Cawley, M.F., \etal 1990, Exp. Astron. 1, 173\\
Comastri, A., \etal 1997, ApJ 480, 534\\
Kataoka, J., \etal 1999, ApJ 514, 138\\
McEnery, J. 1997, Ph.D. Thesis, National University of Ireland\\
Quinn, J., \etal 1999, ApJ 518, (in press)\\
Sikora, M., Madejski, G., Moderski, R., \& Poutanen, J. 1997, ApJ 484, 108\\
Tanaka, Y., Inoue, H., \& Holt, S. S. 1994, PASJ 46, L37\\
Takahashi, T., Madejski, G., \& Kubo, H. 1999, astro-ph/9903099\\
Yaqoob, T., \etal 1997, ApJ 490, L25\\
Yaqoob, T., \etal 1998, ApJ 505, L87\\
\end{document}